\documentclass[aps,prl,reprint,superscriptaddress,showpacs]{revtex4-1}
\usepackage{epsfig}
\usepackage{graphicx}

\begin{document}

\title{Supertransient magnetohydrodynamic turbulence in Keplerian shear flows}

\author{Erico L. Rempel}
\email[]{rempel@ita.br}
\affiliation{Institute of Aeronautical Technology (IEFM/ITA), World Institute
for Space Environment Research (WISER),\\ S\~ao Jos\'e dos Campos -- SP 12228--900, Brazil}
\affiliation{Department of Applied Mathematics and Theoretical Physics (DAMTP), University of Cambridge, Cambridge CB3 0WA, UK}
\author{Geoffroy Lesur}
\author{Michael R. E. Proctor}
\affiliation{Department of Applied Mathematics and Theoretical Physics (DAMTP), University of Cambridge, Cambridge CB3 0WA, UK}

\date{\today}

\begin{abstract}
A subcritical transition to turbulence in magnetized Keplerian shear flows is investigated using a statistical approach. 
Three-dimensional numerical simulations of the shearing box equations with zero net magnetic flux are employed
to determine the transition from decaying to sustained turbulence as a function of the magnetic Reynolds number $Rm$.
The results reveal no clear transition to sustained turbulence as the average lifetime of the transients
grows as an exponential function of $Rm$, in accordance with a type-II supertransient law. 
\end{abstract}

\pacs{47.20.Ft, 47.27.Cn, 47.65.Md, 98.62.Mw}

\maketitle

There has been much progress in recent years on the understanding of the nature of subcritical transitions to turbulence 
in hydrodynamic flows, wherein the laminar state is linearly stable for all Reynolds numbers and
finite amplitude perturbations are necessary to drive the system towards a turbulent state. Numerical and
experimental works indicate that turbulence in systems like plane--Couette and pipe flows is 
a transient phenomenon \cite{couette,faisst04}, although this is still a disputed topic \cite{peixinho06,mellibovsky09}.
Although much has been accomplished for hydrodynamic turbulence in this area, the same cannot be said about
magnetohydrodynamic (MHD) turbulence in systems with subcriticality, for which statistical studies are scarce. 

A classical problem of subcritical transition to turbulence is provided by accretion disks, where, in the
absence of magnetic fields and when shear is the
only stratification, Keplerian flows are linearly stable \cite{biskamp03}.
The rate of mass accretion in the disk depends on the outward transport of angular momentum
and one would think that turbulence is the natural candidate for transport. 
However, both numerical \cite{stone96,hawley99,lesur05}
and laboratory \cite{ji06} experiments have revealed that, if a subcritical hydrodynamic
transition to turbulence was to exist in Keplerian flows, then the associated turbulent transport of angular 
momentum is very weak and unable to match observational constrains.

When a mean vertical field is applied, the magnetorotational instability \cite{balbus91} provides
a linear instability that may result in a strong MHD turbulence capable of generating adequate rates of
angular momentum transport \cite{hawley95,axel95}. If there is no imposed background magnetic 
field, the original (seed) field can decay with time due to finite resistivity.
In that case, some type of nonlinear dynamo process is necessary to amplify and sustain the seed field and,
consequently, the instability in zero net flux simulations is nonlinear \cite{hawley96}. 
Rincon {\it et al.} \cite{rincon07} identified a self-sustaining dynamo 
solution in a magnetized rotating plane--Couette flow in the Keplerian regime, but the steady solutions
were found only in a narrow range of low values of the Reynolds number. Other works have reported
the existence of nonlinear dynamos in turbulent MHD regimes \cite{hawley96},
yet, little is known about how turbulence evolves as a function of the magnetic Prandtl number $Pm$
(the ratio of viscosity to resistivity) and whether it can be sustained 
for long runs or not. 
Recently, Fromang et al. \cite{fromang2} studied the onset of turbulence in accretion disks using an unstratified shearing box with 
zero net magnetic flux and found that turbulence disappears when the magnetic Prandtl 
number falls below a critical value that seems to be a decreasing function of the kinetic Reynolds number. Note,
however, that vertical stratification might change this picture \cite{davis}.

The present letter explores the onset of turbulence in three-dimensional, zero net flux, MHD simulations
of Keplerian shear flows as a function of the magnetic Reynolds number $Rm$. We focus on the statistical aspects
of the subcritical transition, which is an approach that has been unexplored in MHD simulations, 
despite its popularity in the hydrodynamics community. From our results, there
seems to be no clear transition to sustained turbulence. Instead, the average lifetime of the 
turbulence $\tau$ grows as an exponential function of $Rm$, following a type-II
supertransient law \cite{tel08}.

We solve the shearing box equations \cite{goldreich65}, which represent the dynamics in a box moving with the 
disk's angular velocity $\Omega(r_0)$, where $r_0$ is a fiducial radius that determines the location of the center of the box. 
In Cartesian coordinates,
with $\phi \rightarrow y$ and $r \rightarrow r_0 + x$, the shearing box equations read as 

\begin{eqnarray}
\partial_{t}\mathbf{v}+\mathbf{v}\cdot\nabla\mathbf{v} & = & -\frac{\displaystyle 1}{\displaystyle \rho}\nabla P+\frac{\displaystyle 1}{\displaystyle \mu_{0}\rho}\left(\nabla\times\mathbf{B}\right)\times\mathbf{B} \nonumber \\
& & {}-2\mathbf{\mathbf{\Omega\times v}}+2\Omega Sx\hat{\mathbf{x}}+\nu\nabla^{2}\mathbf{v},\label{eq1}\\
\nonumber\\
\partial_{t}\mathbf{B} &=& \nabla\times\left(\mathbf{v\times B}\right)+\eta\nabla^{2}\mathbf{B},\label{eq2}\\
\nonumber\\
\nabla \cdot \mathbf{B} &=& 0, \qquad \nabla \cdot \mathbf{v}=0,
\label{eq3}
\end{eqnarray}
where $\Omega=r^{-3/2}$ for a Keplerian disk, $S=-r\partial_r \Omega = (3/2)\Omega$, $\eta$ is the 
magnetic diffusivity, $\mu_0$ is the magnetic permeability, $\rho$ is the gas density, and $P$ is the pressure. 
The fluid velocity can be decomposed as 
$\mathbf{v} = \mathbf{v}_0 + \mathbf{u}$, where the steady state solution is given by the shear flow $\mathbf{v}_0=-Sx\mathbf{\hat{y}}$
and $\mathbf{u}$ is the perturbation field. 
The equations for $\mathbf{u}$ are solved with the pseudospectral code  
described by Lesur {\it et al.} \cite{lesur05,lesur} using shearing sheet boundary conditions \cite{hawley95}. The $xyz$ aspect ratio
employed is $1 \times \pi \times 1$. The kinetic and magnetic Reynolds numbers are 
defined as $Re=Sd^2/\nu$ and $Rm=Sd^2/\eta$, respectively, where $d$ is the
shearwise size of the box. Time is measured in shear time units, $1/S$. 

\begin{figure}
 \includegraphics[width=7cm]{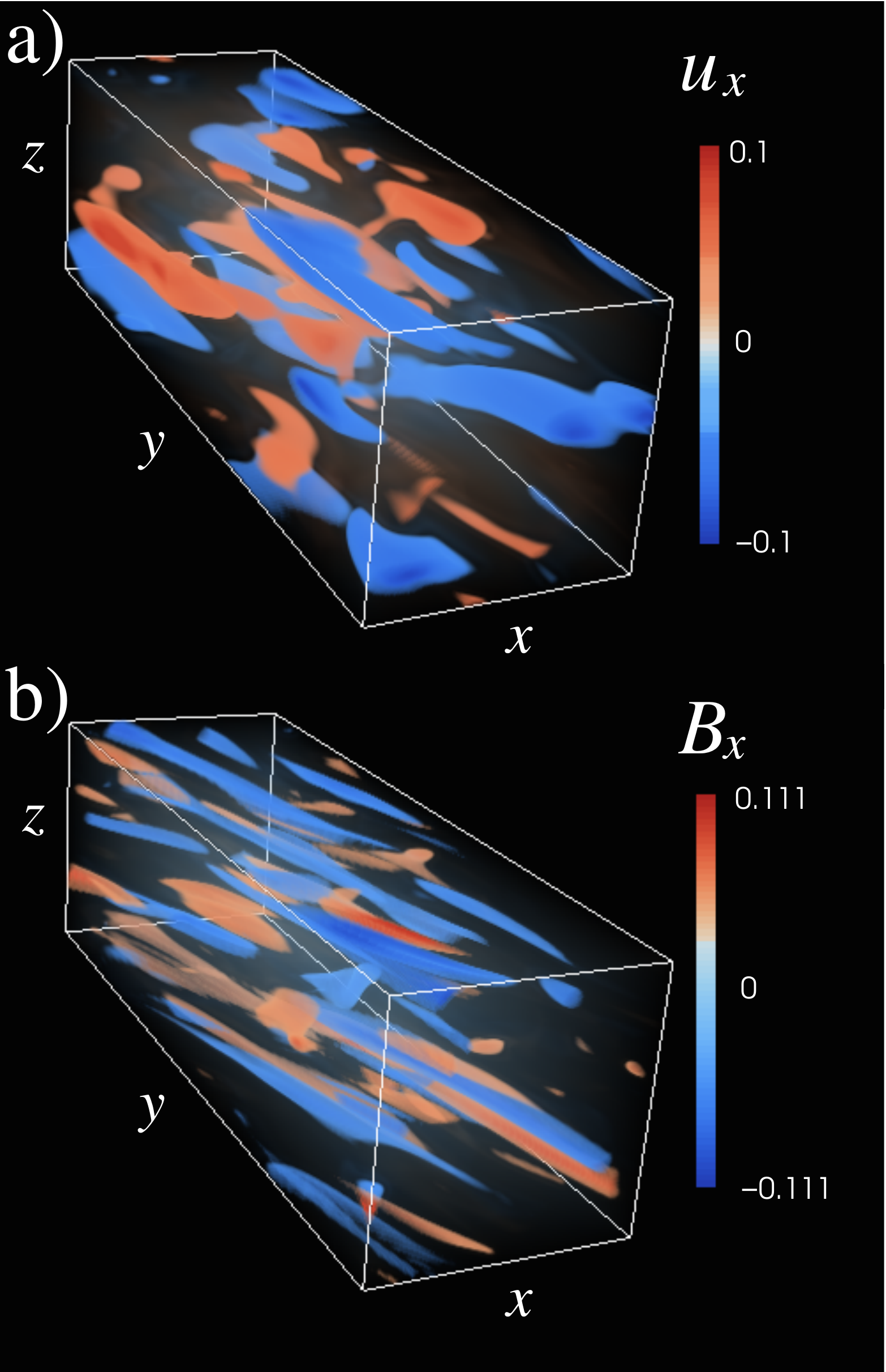}
 \caption{\label{fig patts} (Color online) Volume rendering plots of $u_x$ (a) and $B_x$ (b)
for a turbulent state at $Re=3125$ and $Rm=11000$.}
 \end{figure}

\begin{figure}
 \includegraphics[width=7cm]{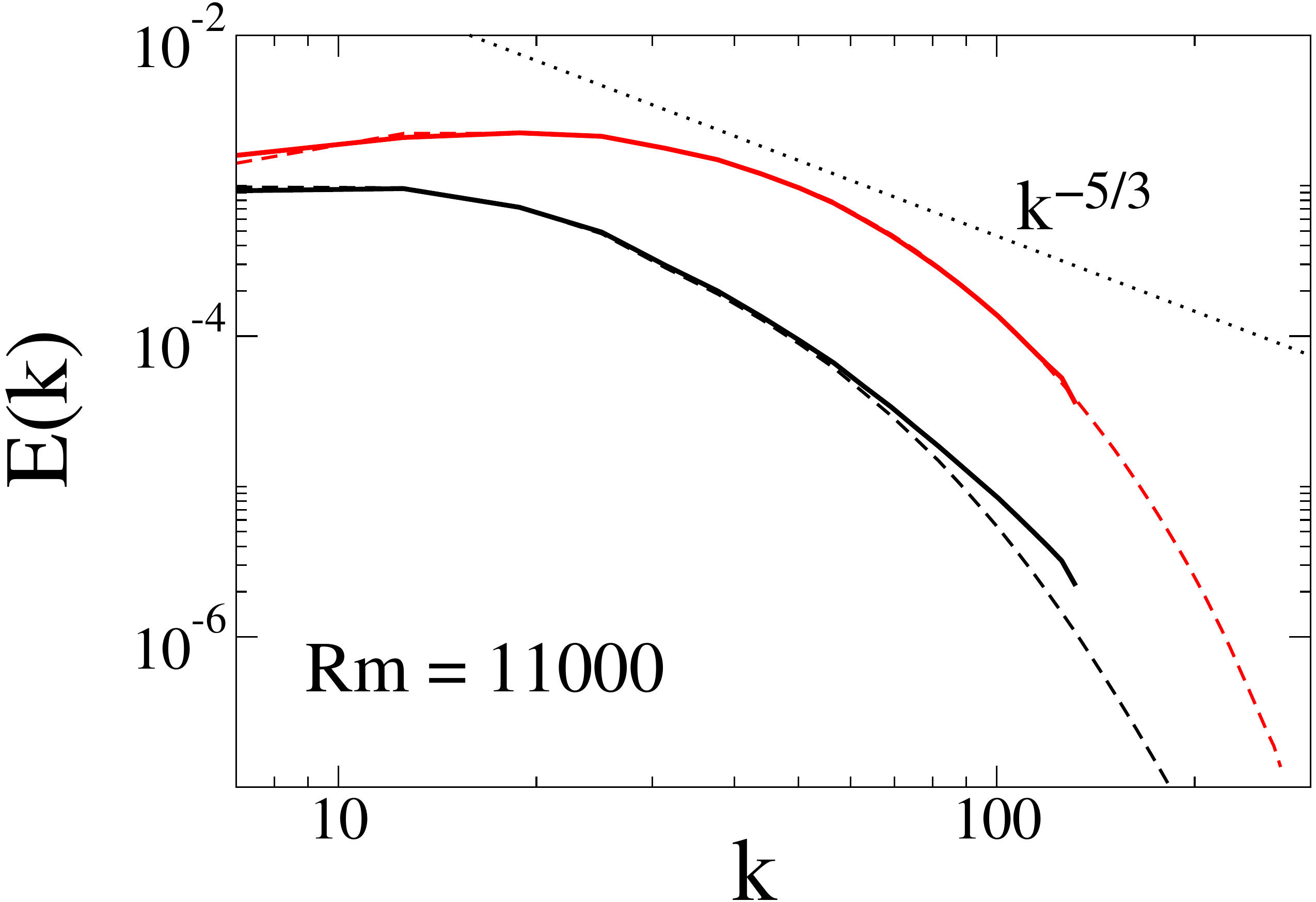}
 \caption{\label{fig spec} (Color online) Time--averaged spectra of kinetic (solid black line) and magnetic 
(solid red line) energies for $Re=3125$ and $Rm=11000$. The dashed lines represent the spectra computed
with doubled resolution.
The $k^{-5/3}$ dotted line is a guide to the eye.}
 \end{figure}

Transition to turbulence is investigated by fixing $Re$ while varying $Rm$. The value of $Re$ is 
taken from Fig. 11 of Fromang {\it et al.} \cite{fromang2}, where for $Re=3125$ there is
a transition to turbulence somewhere between $Rm=6250$ and $Rm=12500$, thus, the magnetic Prandtl number $Pm = Rm / Re > 1$. 
Since we are not interested in the kinematic dynamo regime, the initial growth of 
a small seed magnetic field is not important to our discussions. Our question is: given a 
turbulent state, can it be sustained for all time? If not, how long it takes, on average, to decay? 
The turbulent states are obtained from long runs where the velocity and magnetic fields are 
initialized with a large--scale random noise (only the four largest modes are nonzero at $t=0$) with amplitude of the
order of 1. After some time, energy spreads to other modes and the random fields settle to a turbulent state.
Figure \ref{fig patts} illustrates the typical turbulent patterns found at $Re=3125$
and $Rm=11000$ for the radial component of the velocity (upper panel) and magnetic (lower panel) 
fields at $t=500$. Since $Pm > 1$ and there is no proper large scale dynamo, 
the characteristic spatial scales in the magnetic fields are smaller than in the velocity
fields. The numerical resolution employed is $64 \times 128 \times 64$, which is the same resolution used for this problem in Ref. \cite{fromang2}.
The convergence of the code is checked by comparing the time-averaged power spectra obtained with this
run with a higher--resolution run ($128 \times 256 \times 128$). The results are shown in Fig. \ref{fig spec}, where
the solid black line represents the lower--resolution kinetic energy spectrum and the solid red line the lower--resolution magnetic energy spectrum.
The higher--resolution spectra are plotted in dashed lines and agree quite well with the lower resolution runs. There is a disagreement 
in the small scales in the kinetic spectra, but this has a low impact on the global behaviour. Table 1 contains
time--averaged values for the kinetic and magnetic energies, 
as well as for the Reynolds and Maxwell stresses,
which are defined, respectively, as $\alpha_{Rey}=\left<u_xu_y\right>$
and $\alpha_{Max}=-\left<B_xB_y\right>$.  The quantities are computed with both resolutions for $Re=12500$.
In the remainder of this letter, we adopt the lower resolution.
Note that the lack of a properly defined inertial range in the power spectra indicates that the turbulence is not fully developed.

\begin{table}[h]
\centering
\begin{minipage}{\columnwidth}
\caption{Time-averaged quantities for $Rm=12500$.}
\begin{tabular}{|c|c|c|}
\hline 
Res. & $64\times 128 \times 64$ & $128 \times 256 \times 128$ \tabularnewline
\hline
\hline 
$\left<E_{kin}\right>_t$ & $2.11 \times 10^{-3}$ & $2.07 \times 10^{-3}$ \\
$\left<E_{mag}\right>_t$ & $8.45 \times 10^{-3}$ & $8.52 \times 10^{-3}$ \tabularnewline
$\left<\alpha_{Rey}\right>_t$ & $5.22 \times 10^{-4}$ & $5.24 \times 10^{-4}$ \tabularnewline
$\left<\alpha_{Max}\right>_t$ & $3.68 \times 10^{-3}$ & $3.69 \times 10^{-3}$ \tabularnewline
\hline
\end{tabular}
\end{minipage}
\end{table}

\begin{figure}
 \includegraphics[width=7cm]{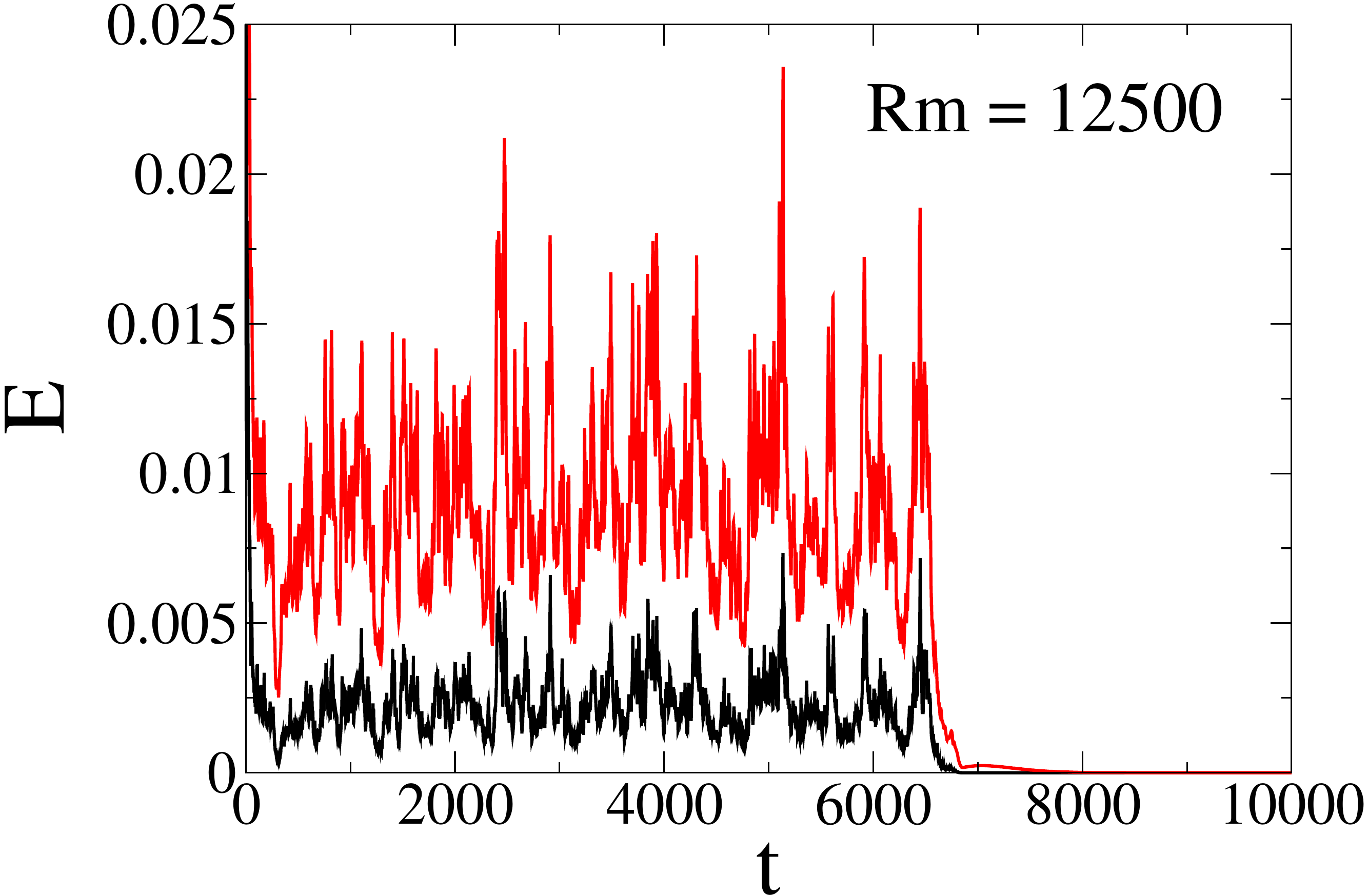}
 \caption{\label{fig ts} (Color online) Time series of the magnetic (red) and kinetic (black) energies
showing transient turbulence for $Re=3125$ and $Rm=12500$.}
 \end{figure}

The turbulence shown in Fig. \ref{fig patts} eventually decays toward the laminar state. The decay
time is a function of the initial perturbation and $Rm$.
Figure \ref{fig ts} shows an example of a time series of the magnetic (red) and kinetic (black) energies
for $Re=3125$ and $Rm=12500$. After more than 6000 time units the turbulence suddenly
decays and the system converges to the laminar state.

\begin{figure}
 \includegraphics[width=7cm]{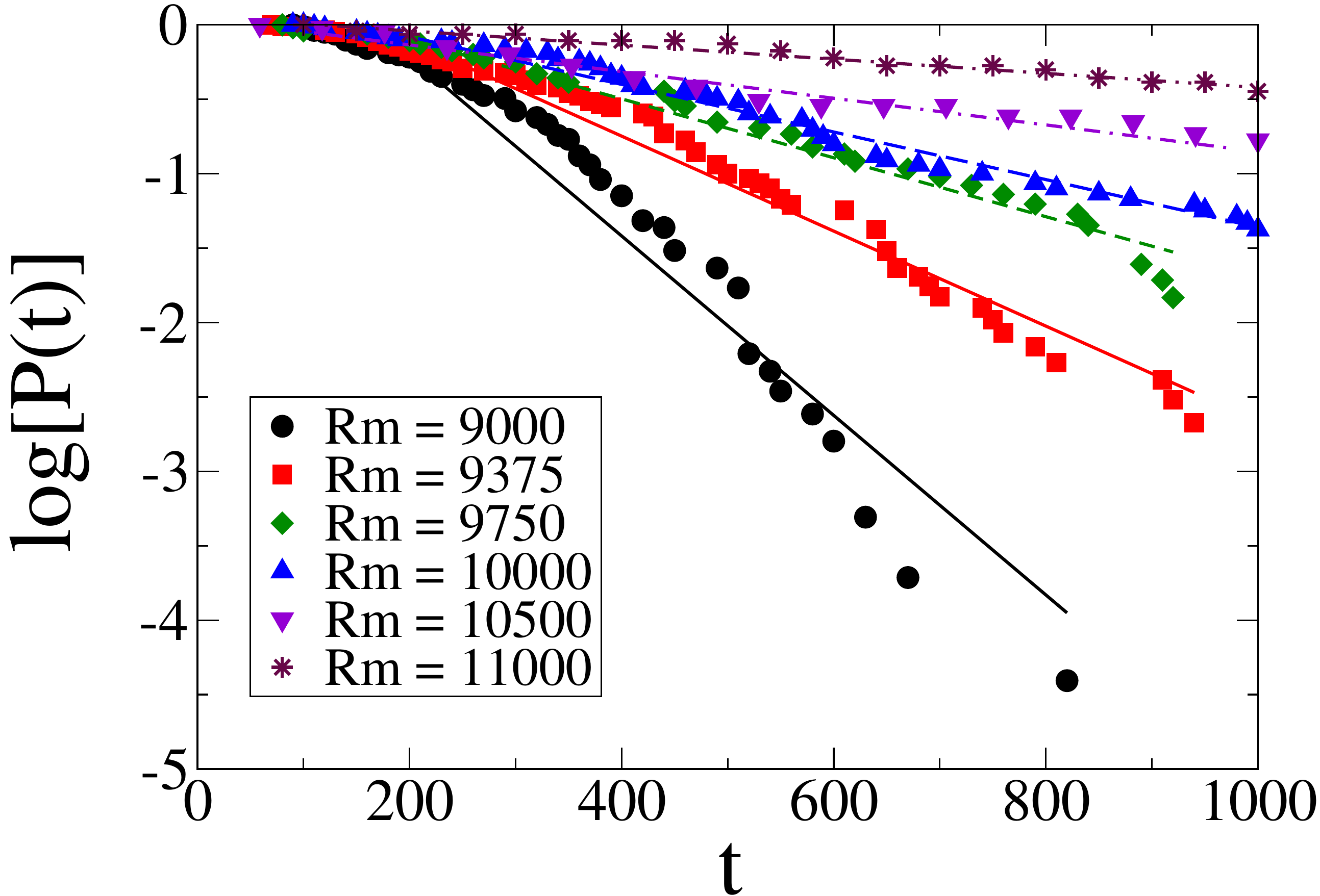}
 \caption{\label{fig laws} (Color online) The logarithm of $P(t)$, the probability of turbulent lifetime
$\geq t$, as a function of $t$ for $Re=3125$ and six different values of $Rm$;
the straight lines are fitted with $P(t,Rm) = \exp[-t/\tau(Rm)]$, where $\tau$ is the characteristic turbulence lifetime.}
 \end{figure}

The high variability of the turbulence decay times as a function of the initial condition for a given $Rm$ suggests
the presence of a chaotic saddle in the phase space \cite{faisst04,skufca06}. The role of chaotic
saddles in generating chaotic transients in dynamical systems is well known and, as mentioned in previous 
studies on spatiotemporal systems \cite{rempel05}, the average lifetime of the chaotic transients grows as a function 
of the fractal dimension of the stable manifold of the chaotic saddle.
In order to obtain the average lifetime $\tau(Rm)$ of the turbulent transients we compute 
$P(t)$, the probability of finding a turbulent state at time $t$. 
For each value of $Rm$, $P(t)$ is obtained from a set of more than 50 and up to 90 different simulations. Each 
simulation has a turbulent state as initial condition and the simulation stops when either $t=1000$ or
the laminar state is reached, generating a data set of relaminarization times.
The turbulent initial conditions are taken from a long run at $Rm=13000$. 
To ensure that the initial conditions represent uncorrelated states,
they were selected from the referred long turbulent run with 500 shear--time units apart from each other.
Figure \ref{fig laws} shows
the logarithm of $P(t)$ as a function of $t$ for $Re=3125$ and a range of different values of $Rm$. The straight
lines are least square fits that represent the relation $P(t,Rm) = \exp[-t/\tau(Rm)]$, which is expected for transients 
due to chaotic saddles \cite{hsu88,faisst04}.
The values of $1/\tau$ are readily obtained from the slopes of the fitted lines.
As $Rm$ increases, the slopes decrease, reflecting longer decay times.
The convergence is not very good for the lowest values 
of $Rm$, probably due to a lower fractal dimension of the stable manifold of the chaotic saddle, which implies
that a much higher number of initial conditions would be necessary to obtain better statistics. 

\begin{figure}
 \includegraphics[width=7cm]{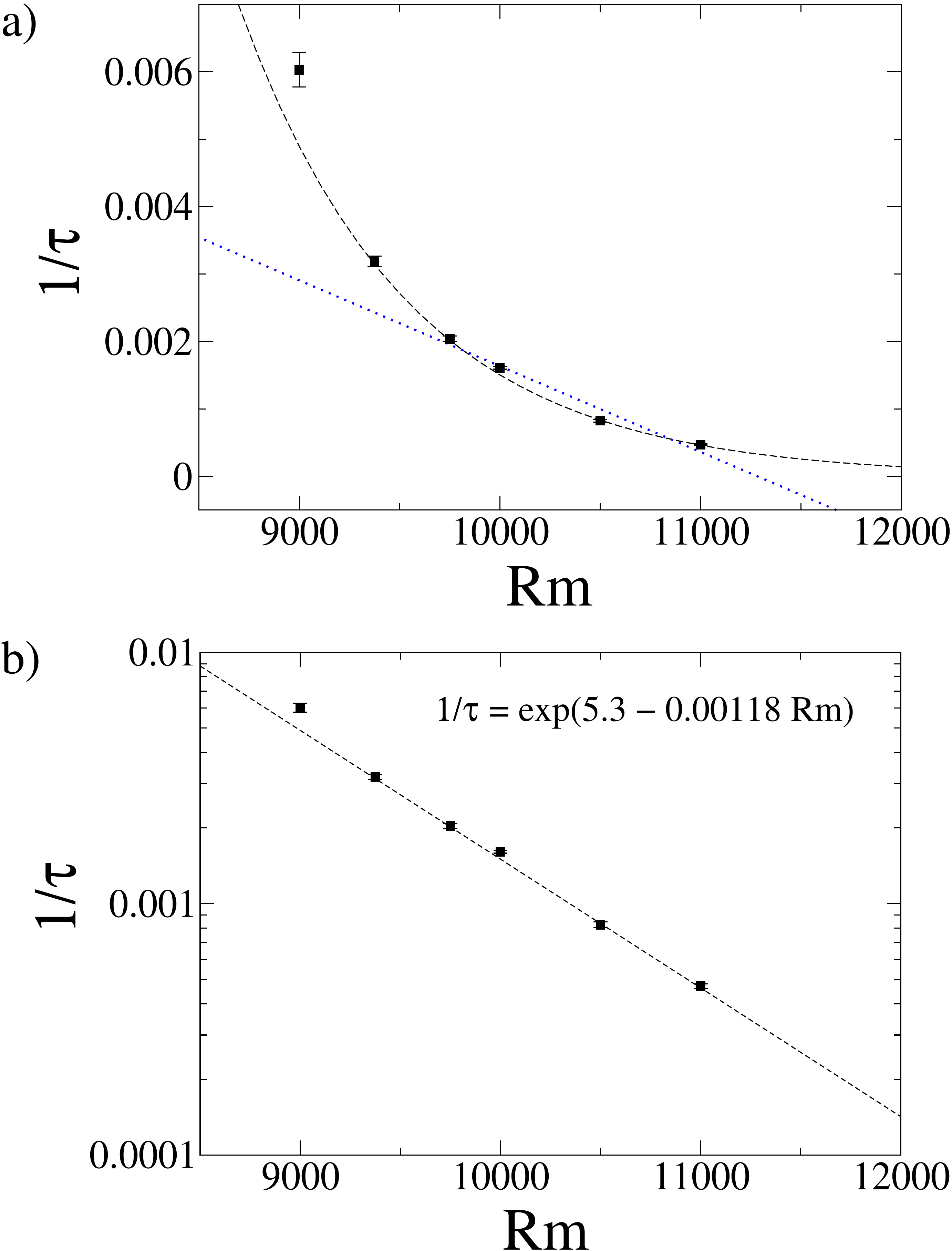}
 \caption{\label{fig taus} (Color online) The characteristic turbulent lifetimes as a function of $Rm$, obtained
from fits to the data shown in Fig. \ref{fig laws}; (a) linear scales, the dashed curve 
is an exponential fit and the dotted straight line is a linear fit using only the last
four points; (b) same data as in (a) but in log--lin scales, 
with the exponential fit.}
 \end{figure}

A graph of $\tau$ versus $Rm$ is plotted in Fig. \ref{fig taus}. The top panel shows the data in linear scales
and reveals that the inverse lifetime can be well fitted with an exponential law as a function of $Rm$. The
dashed curve is given by $1/\tau = \exp(5.3-0.00118 Rm)$. The lower panel shows the same data in log--lin
scales. These results indicate that the turbulence lifetime follows a type-II supertransient law \cite{tel08}
as a function of $Rm$. In this case, there is no apparent transition to sustained turbulence, and the laminar
state is the global attractor.
If there were a critical $Rm$ for transition to sustained turbulence via a crisis
mechanism, a linear decay of $1/\tau$ with $Rm$ would have been expected \cite{peixinho06}. The results for a linear
fit using only the last four values of $Rm$ are given by the dotted blue line in Fig. \ref{fig taus}(a),
which represents $1/\tau=0.0144-1.273\times 10^{-6} Rm$. This fit provides a critical point $Rm_c \sim 11312$,
which is clearly incorrect, since Fig. \ref{fig ts} shows that even at $Rm = 12500$ the turbulence
is still decaying. 
 
Note that the high values of $Re$ and $Rm$ shown in this letter are based on the shear rate $S$. 
If the Reynolds numbers are computed based on the root mean square velocity, one obtains a maximum of $Re\sim 180$ and $Rm\sim 650$.
Thus, it has been shown that, at least for intermediate Reynolds numbers and magnetic Prandtl numbers greater than one, MHD turbulence
in the incompressible shearing box equations with zero net magnetic flux is not self-sustained. The average decay time $\tau$ follows a 
type-II supertransient law,
with an exponential increase of $\tau$ as a function of the magnetic Reynolds number. This is in accordance with
results found for subcritical transitions to turbulence in hydrodynamic flows such as the pipe flow \cite{faisst04} and plane Couette flow \cite{couette},
where turbulence seems to be always transient. 
Supposing that our conclusions can be extended to realistic values of the Reynolds numbers, this suggests that the transient dynamics is much more 
relevant to the accretion problem than the final attractor.

\begin{acknowledgments}
 ELR thanks DAMTP for the hospitality. ELR acknowledges support by CNPq (Brazil). GL acknowledges support by STFC. The simulations presented in this paper were performed using the cluster for high-performance computing of the National 
Institute for Space Research (LAC/INPE, Brazil) and the Darwin Supercomputer of the University of Cambridge.
\end{acknowledgments}

\end{document}